\documentclass[article]{aa} 

\graphicspath{{figures/}}

\usepackage{comment}

\begin{document}

\title{Gamma radiation from cosmic rays escaping a young supernova remnant: The case of Cas A}
\titlerunning{Gamma rays from cosmic rays escaping Cas A}
\authorrunning{Blasi, P.}

\author{Pasquale Blasi$^{1,2}$}

\institute{Gran Sasso Science Institute (GSSI), Viale Francesco Crispi 7, 67100 L’Aquila, Italy \and INFN-Laboratori Nazionali del Gran Sasso (LNGS), via G. Acitelli 22, 67100 Assergi (AQ), Italy}

\date{\today}

\abstract{The escape of accelerated particles from supernova remnants remains one of the central and yet least understood aspects of the origin of cosmic rays. Here we use the results of the recent LHAASO observation of gamma rays from a region of $\sim 2$ degrees around the Cassiopeia A (Cas A) supernova remnant to constrain the process of particle escape from this remnant. We dedicate special attention to discuss the impact of shock evolution and particle propagation in the near source region on the gamma ray emission. This is very important to assess the possibility that very high-energy protons may have been accelerated in past activity of Cas A. Given the core collapse nature of Cas A and its young age ($\sim 340$ years), the non detection of $\gtrsim 100$ TeV gamma rays from this remnant allows us to draw some conclusions on the role of supernova explosions for the origin of cosmic rays at the knee. In particular we calculate the spectrum of cosmic rays that may have escaped this remnant to make a clear distinction between the instantaneous maximum energy and the one that appears as a flux suppression in the time integrated spectrum from an individual supernova. LHAASO observation of the region around Cas A as well as gamma ray observations of the remnant itself confirm that this remnant is not operating as a PeVatron and that, although in the early stages the maximum energy may have reached $\sim$Peta-electronVolt, the number of such particles in the surroundings of Cas A is exceedingly small.} 


\maketitle

\section{Introduction}
From the point of view of energetics, supernova remnants (SNRs) are hard to beat as potential sources of the bulk of Galactic cosmic rays (CRs): an efficiency of $\sim 3-10\%$ is required in order to reproduce the spectrum of CRs observed at Earth \cite[]{Blasi2013,Amato2018,Blasi2019}. While this conclusion is rather solid in terms of the bulk of CRs in the Galaxy, it is all but clear which sources contribute the most to the region of the knee ($\sim 3\times 10^{15} eV$). At present no SNR has been detected in gamma rays with a cutoff or flux suppression starting at $E\gtrsim 100$ TeV, which would indicate effective acceleration of CRs to Peta-electronVolt energies. Nevertheless, a clearly negative conclusion in this respect would require caution: first, the highest energies are expected to be reached at early times in the SNR evolution, and so far we have not been lucky enough to catch one such events in gamma rays. Second, SNRs come in many flavors and occur in different environments: it cannot be excluded that in some such environments the maximum energy of accelerated particles may approach $\sim$few PeV values. Third, our conclusion that SNRs are unlikely to be PeVatrons is partly due to the lack of observational evidence of $\gtrsim 100$ TeV gamma rays from known SNRs, and partly due to the fact that streaming instability is unable to warrant such high energies, both in its resonant \cite[]{Lagage1,Lagage2} and non-resonant \cite[]{Bell:2004p737} versions. This last point is the weakest, in that one could argue that other types of instabilities or other processes may lead to higher maximum energies. It has been proposed that this may indeed happen when magnetic field is amplified due to the excitation of acoustic instabilities \cite[]{Beresnyak2009,Drury2012} or if particle scattering is mediated by mirroring \cite[]{Bell2025}. In the latter article, the case of Cas A was considered as a possible instance of an accelerator where $\gtrsim$Peta-electronVolt energies could be achieved due to mirroring. 
 
At this point it is of the utmost importance to define the maximum energy of accelerated particles in a clear and agreed upon manner: at each time $t$ the maximum energy of accelerated particles has a given value, which can in principle be as high as Peta-electronVolt or even larger in some cases. However, for typical parameters of a SNR, this happens in very early stages of the evolution, when the number of particles accelerated at such high energies is exceedingly small and the corresponding spectrum of CRs liberated into the interstellar medium (ISM) is correspondingly steep. The maximum energy that is relevant from the point of view of assessing the role of SNRs as potential PeVatrons is the maximum energy at a time where most of the SNR energy is being processed. This happens at the beginning of the Sedov-Taylor phase \cite[]{Schure-Bell:2013,Schure:2012p3068,martina,pierre,pierre2021}: above such energy the spectrum of CRs liberated into the ISM steepens considerably with respect to the standard spectrum expected based on diffusive shock acceleration (DSA), although there is no exponential suppression. In order for SNRs to act as PeVatrons for Galactic CRs, the maximum energy of accelerated particles at the beginning of the Sedov-Taylor phase of the SNR must be $\sim 1-3$ PeV. 

The process of particle acceleration is intimately connected to that of magnetic field amplification, which in turn is shaped by the escape of particles from the acceleration region. The present understanding of the escape of CRs from SNRs is that it happens in two stages: while particle acceleration is always at work at the forward shock, the maximum energy drops with time and at any given time particles at the maximum energy reached at time $t$, $E_{\rm max}(t)$, can escape the system from the upstream region. This escape process is at work during the ejecta dominated and the Sedov-Taylor phases. By the end of the Sedov phase, particles with energy higher than the maximum energy achieved at the end of the Sedov phase (typically of order $\sim$TeV) will have escaped the remnant, while lower energy particles have been advected downstream of the shock and continuously lose energy adiabatically (for protons). These particles will eventually leave the remnant at some point during the radiative phase. The spectrum that overall is contributed by an individual SNR is the sum of these two contributions, which is typically not a perfect power law \cite[]{pierre,pierre2021}.

With the development of more sensitive large area high energy gamma ray telescopes, such as LHAASO, it has become possible and in fact desirable to look for signatures of the gamma ray emission due to CRs that escaped a young SNR in the past. This is especially promising for the Cas A SNR, which is about $340$ years old and resulted from a SN explosion taking place in the wind of the pre-supernova red giant star, an occurrence that has been proposed as possibly leading to acceleration to Peta-electronVolt energies \cite[]{Schure-Bell:2013}. 

LHAASO has observed a region of $\sim 2$ degrees around Cas A \cite[]{Cao2024}, to test the possibility that very high-energy protons, accelerated during the early stages of the evolution of this SNR may produce gamma rays today. This observation has led to no detection, which allowed the collaboration to infer upper limits on gamma rays with $10\leq E_\gamma\leq 1000$ TeV from the region around Cas A. 

Here we will discuss what the implications are of such limits for the models of escape of CRs from Cas A and for the maximum energy of the particles accelerated at this remnant. The article is structured as follows: in Sec. \ref{sec:escape} we will discuss the intimate connection between particle escape from the acceleration region and the maximum energy achieved at time $t$, in the case of SNe exploding in the wind of the pre-supernova star. In Sec. \ref{sec:transport} we will discuss the expected type of cosmic ray transport in the region around the source and its implications in terms of average column density traversed by particles while escaping, specializing the discussion to the case of Cas A. In Sec. \ref{sec:gamma} we will show the results of calculations of the gamma ray emission from the region around Cas A for different cases of SN shock evolution compatible with the present Cas A forward shock measured properties. We will conclude and discuss the prospects of this type of observations in Sec. \ref{sec:discuss}.

\section{Escape from a core collapse SNR}
\label{sec:escape}

This section describes our current understanding of the escape of CRs from a core collapse SNR associated with a SN which occurred in the wind of the pre-supernova red giant star. The discussion is general but specialized to the parameters of Cas A, which is thought to be the remnant of a type IIb supernova explosion \cite[]{Krause2008,Rest2011}. It is located at a distance of 3.4 kpc from the Sun \cite[]{Reed1995} which puts the forward shock at a distance of $\sim 2.8$ pc from the explosion center at the present time \cite[]{Vink2022}. The mass of the ejecta associated with the supernova explosion has been estimated to be $M_{ej}=2-4~M_\odot$ (see \cite{Vink2022} and references therein). 

The current velocity of the forward shock has been measured to be $\sim 5700-5800$ km/s \cite[]{Vink2022}, that helps nailing down the overall dynamical history of the remnant. The shock is expected to propagate in the wind of the pre-supernova red giant star, with density as a function of the radial coordinate $r$:
\begin{equation}
    \rho_w(r)=\frac{\dot M}{4\pi r^2 v_w},
\label{eq:windprofile}
\end{equation}
where $\dot M=\dot M_{-5}\times 10^{-5}\rm M_\odot/yr$ is the rate of mass loss and $v_w=v_6 \times 10$ km/s is the velocity of the wind. Both quantities are normalized to typical values of these quantities. The wind excavates a bubble in the surrounding medium up to a distance that is set by the pressure balance between the wind and the external ISM. Assuming $n_H=1~\rm cm^{-3}$ and $T_H=10^4~K$ for the density and temperature of the ISM, the size of the bubble $R_b$ is expected to be determined by
$$
\rho_w v_w^2 \approx n_H k_b T_H \to R_b\approx \left( \frac{\dot M v_w}{4\pi n_H k_B T_H} \right)^{1/2} \approx    
$$
\begin{equation}
\hskip .8cm    \approx 2~\rm pc~ \dot M_{-5}^{1/2} v_6^{1/2} \left( \frac{n_H}{1\rm cm^{-3}}\right)^{-1/2}\left( \frac{T_H}{10^4\rm K}\right)^{-1/2}.
\label{eq:cavity}
\end{equation}
This simple and rough estimate shows that the size of the bubble is comparable with the present shock location. However, the measurements of the time dependence of the shock speed carried out by \cite{Vink2022} show no evidence of the shock crossing the edge of the cavity, being consistent with shock propagation in a $\sim 1/r^2$ density profile. Requiring that the cavity is larger than 2.8 pc, implies $\dot M_{-5} v_6 \gtrsim 2$, which is compatible with the values of parameters often adopted for Cas A: $\dot M_{-5}=1.5$ and $v_6=2$ (see for instance \cite{zira2014}). Since Eq. \ref{eq:cavity} is only a rough estimate, we will retain some flexibility in choosing the numerical values of the quantities involved. In any case, it is clear that the shock is approaching the edge of this cavity. In addition, one can estimate the position of the shock when the mass of the swept up gas equals the mass of the ejecta, which corresponds to a rough beginning of the Sedov phase: 
\begin{equation}
    R_{ST}\approx \frac{M_{ej} v_w}{\dot M} \approx 2~\rm pc ~ \dot M_{-5}^{-1} v_6 \left( \frac{M_{ej}}{2 M_\odot}\right).
    \label{eq:ST}
\end{equation}
It follows that Cas A is also expected to be close to the beginning of the Sedov-Taylor phase, a conclusion confirmed by the more refined calculation of the shock motion presented below.

The motion of the forward shock in the wind region is described here using the approach put forward by \cite{TangChevalier}, that we summarize below. Let us first introduce a characteristic time and radius as
\begin{eqnarray}
t_{ch}=E_{SN}^{-1/2}M_{ej}^{3/2}\eta_s^{-1}=1772~\rm yr ~E_{51}^{-1/2} \left(\frac{M_{ej}}{M_\odot}\right)^{3/2}\dot M_{-5}^{-1}v_6\\
R_{ch}=M_{ej}\eta_s^{-1}=12.9~\rm pc~\left(\frac{M_{ej}}{M_\odot}\right)\dot M_{-5}^{-1}v_6,
\end{eqnarray}
where $\eta_s=\dot M/4\pi v_w$ and $E_{51}=E_{SN}/10^{51}\rm erg$. The motion of the forward shock is parametrized in terms of the dimensionless quantities $t_*=t/t_{ch}$ and $R_*=R_s/R_{ch}$:
\begin{equation}
    R_*(t_*) = \left[ \left(\zeta t_*^{\frac{k-3}{k-2}} \right)^{-s} + \left(\xi t_*^{2} \right)^{-s/3}\right]^{-1/s},~~~~~k>5,
    \label{eq:Radius}
\end{equation}
where $s$ is a smoothing parameter, $k$ describes the drop in density of the profile of the ejecta (below we consider the two cases $k=9$ and $k=12$, which comprise the level of uncertainty in this parameter for a core collapse supernova). The parameters $\zeta$ and $s$ were fitted to the results of simulations by \cite{TangChevalier} for the case of expansion of the supernova shell into the wind of the pre-supernova star, with density profile as in Eq. \ref{eq:windprofile}, and are listed in Table \ref{table:t1}, while $\xi=3/2\pi$. The shock velocity can be easily calculated as 
\begin{equation}
v_s(t)=\frac{d R_s(t)}{dt}= \frac{R_{ch}}{t_{ch}}v_*(t_*),
\label{eq:velocity}
\end{equation}
where $v_*(t_*)=\frac{d R_*(t_*)}{d t_*}$. Eq. \ref{eq:Radius} parametrizes the transition from the ejecta dominated stage to the Sedov-Taylor phase, that can be shown to occur at the following transition time and radius:
\begin{equation}
    t_{*,tr}=\left( \frac{\xi}{\zeta^3}\right)^{\frac{k-2}{k-5}},~~~~~R_{*,tr}=\frac{\zeta}{2^{1/s}}\left( \frac{\xi}{\zeta^3}\right)^{\frac{k-3}{k-5}}.
\end{equation}
It is worth noticing how the transition occurs at a radius (and time) that are substantially different from those that one might estimate in a naive way. For instance, for $k=9$ ($k=12$) one can see that the transition to the Sedov phase occurs at a distance from the explosion center that is $\sim 4 R_{ST}$ ($\sim 2.78R_{ST}$), where $R_{ST}$ is the rough estimate in Eq. \ref{eq:ST}. In practice, this means that the transition from the ejecta dominated to the Sedov phase is, in general, rather smooth and it may be difficult to pinpoint an exact time or radius to adopt as a benchmark value. Moreover, these quantities depend on the density profile of the ejecta (parameter $k$ above), which is not well constrained. This uncertainty reflects on the value of the maximum energy reached by the particles at the beginning of the Sedov phase, as discussed below. 

Now that the dynamics of the forward shock is well defined, we discuss the escape of the accelerated particles from the forward shock, a phenomenon that is tightly connected to the process of magnetic field amplification \cite[]{Schure-Bell:2013,martina,pierre}. The differential spectrum of particles accelerated at the shock is parametrized here as $n(E,t)=A(t)\left( E/E_0 \right)^{-2-\beta}$ (for $E<E_{max}(t)$), where $E_0=1$ GeV and the normalization constant $A$ is determined by requiring that accelerated particles carry a fraction $\xi_{CR}$ of the ram pressure at the shock location:
\begin{equation}
\frac{1}{3}\int_{E_0}^{E_{max}(t)} dE~ E ~n(E,t) = \xi_{CR} \rho_w v_s^2,
\end{equation}
which returns $A=3\xi_{CR}\rho_w v_s^2/E_0^2 I(\beta)$, where:
\begin{equation}
    I(\beta)=
\begin{cases}
  \ln\left(\frac{E_{max}}{E_0}\right), & \text{if } \beta = 0 \\
  \frac{1}{\beta}\left[ 1 - \left(\frac{E_{max}}{E_0}\right)^{-\beta}\right], & \text{if } \beta > 0.
\end{cases}
\end{equation}
Here we explicitly assumed that the SN shock propagates in the density profile $\rho_w (r)$ of the wind of the pre-supernova star and $v_s$ is given by Eq. \ref{eq:velocity}. The local density is $\rho_w(r=R_s)$, calculated at the position of the shock at time $t$.

A few comments are in order: 1) the spectrum of accelerated particles at a strong shock, as the one expected in the case of a Type IIb SNR, is $\propto p^{-4}$ ($E^{-2}$ for relativistic particles), since the Mach number is very large. Here we allowed for the possibility to have a slightly steeper spectrum ($\beta>0$). This is motivated by the fact that, as discussed by \cite{Haggerty2020,Caprioli+2020}, in the presence of magnetic field amplification, magnetic perturbations move in the downstream at basically the Alfv\'en speed in the amplified field, which results in steeper spectra of accelerated particles. 2) The level of magnetic field amplification is determined by the current in the form of escaping CRs \cite[]{Bell:2004p737}, which is however sensitive to the maximum energy, which in turn is shaped by magnetic field amplification. This implies that the normalization of the spectrum needs to be calculated by iterations, since $I(\beta)$ depends weakly on $E_{max}$. 3) The normalization of the spectrum of accelerated particles using the pressure implies that $\xi_{CR}$ used here corresponds to about 1/3 of the efficiency that is typically quoted in phenomenological approaches to CR acceleration (the typical 10\% corresponds in our formalism to $\sim 3\%$). 4) The efficiency of particle acceleration adopted here is small enough that one can, in first approximation, neglect the dynamical shock modifications induced by CR pressure and escape, as also discussed by \cite{Caprioli+2020}. This level of efficiency is compatible with that required based on the CR flux measured at the Earth \cite[]{pierre}.

One might argue that the parameters of particle acceleration may be tuned to the direct observation of gamma rays from the remnant itself (see \cite{Cao2024} and references therein for joint detection of gamma rays by Fermi-LAT and LHAASO). However, this is not necessarily the optimal way to proceed for a variety of reasons: first, the interpretation of the observed gamma ray emission in terms of inelastic proton scatterings in the wind leads to requiring exceedingly large acceleration efficiency, $\gtrsim 30\%$. Such large efficiencies would however lead to large magnetic field, which in turn would produce steep CR spectra, plausibly making the flux of CRs in the high energy region smaller (effect of the postcursor), thereby contradicting a hadronic interpretation of the gamma ray emission. The effect of the magnetic postcursor \cite[]{Caprioli+2020} has not been included in existing calculations and will be the subject of a dedicated investigation. Second, as a consequence of the large efficiency required by a hadronic interpretation, the contribution of inverse Compton scattering of electrons has been considered \cite[]{zira2014,Veritas2020,Cao2025}, leading to a potentially good fit to observations. However, it is not clear if in this class of models the radio observations are well described, given the fact that part of this emission seems to come from the region of the reverse shock \cite[]{CasAradio1995}. In fact, evidence of particle acceleration at the reverse shock also exists in the hard X-ray band \cite[]{Grefenstette2015,Woo2025}. Measurements of the gamma ray emission due to escaping particles are insensitive to the leptonic component, since the large magnetic field at the shock makes high energy electrons lose their energy in the shock region, leaving only high-energy protons free to escape from the upstream region. 

Given the large uncertainties in fixing $\xi_{CR}$ from direct gamma ray observations of the SNR, here we use an educated guess on the value of this parameter and we make use of observations of the diffuse emission from the region around Cas A as a tool to constrain the acceleration efficiency. The number density of particles inside the acceleration region at time $t$ can be written as:
\begin{equation}
    n(>E,t)=\frac{3\xi_{CR} \rho_w(R_s) v_s^2}{(\beta+1) E_0 I(\beta)} \left( \frac{E}{E_0} \right)^{-\beta-1} \,\,\,\,\,E<E_{max}(t),
\end{equation}
corresponding to an electric current of escaping protons at the maximum energy:
\begin{equation}
    J(>E,t)=e n(>E,t) v_s = \frac{3 e \xi_{CR} \rho_w(R_s) v_s^3}{(\beta+1) E_0 I(\beta)} \left( \frac{E}{E_0} \right)^{-\beta-1},
\end{equation}
limited to particles with energy $E\sim E_{max}(t)$ at each given time. 

If the preexisting magnetic field $B_0(r)$ at the position $r$ is such that the condition $J(>E)E (v_s/c)\geq B_0^2/4\pi$ is fulfilled, a non resonant hybrid instability in excited \cite[]{Bell:2004p737}, with a growth rate $\gamma_{CR}=k_{CR}v_A$, where $k_{CR}=4\pi J(>E)/c B_0$ is the wavenumber at which the growth is fastest, and $v_A=B_0/\sqrt{4\pi \rho_w}$ is the Alfv\'en speed in the undisturbed wind. As discussed in much of previous literature, the maximum energy of the accelerated particles is determined by the condition that the instability develops for at least a few e-folds (following \cite{Bell:2004p737,martina,pierre}, here we assumed that 5 e-folds are sufficient, while a larger number results in lower values of the maximum energy). This condition results in the following value of $E_{max}(t)$ as a function of time:
\begin{equation}
    \frac{E_{max}(t)}{E_0}=\left[ \frac{3 e \xi_{CR} v_s^3}{5 (\beta+1) E_0 I(\beta) c} \sqrt{\frac{\dot M}{R_s^2 v_w}} t\right]^{\frac{1}{1+\beta}},
    \label{eq:Emax}
\end{equation}
where we stress again that $I(\beta)$ depends on $E_{max}$, though weakly, so that this is an implicit equation for the maximum energy. 

At each given time $t$ and within a time range $dt$, only particles with energy in a range $dE_{max}$ around $E_{max}(t)$ can escape. This means that the spectrum of particles escaping the remnant can be written as 
\begin{eqnarray}
    N(E) dE = n(>E) v_s 4\pi R_s^2 dt|_{E=E_{max}(t)} \to\\ N(E) = n(>E) v_s 4\pi R_s^2 \frac{dt}{d{E_{max}}(t)},
    \label{eq:spectrum}
\end{eqnarray}
where $dE_{max}/dt$ can be calculated from Eq. \ref{eq:Emax}. It is worth stressing that for a SNR of age $T_{SN}$, only the particles with energy $\gtrsim E_{max}(T_{SN})$ had a chance to escape the remnant, while lower energy particles stay confined downstream until the shock slows down. For a SNR such as Cas A, $T_{SN}=340$ years, and only very high-energy particles could escape (see below for a more quantitative assessment) and can produce gamma rays in the surroundings of the SNR.  

As pointed out earlier, Cas A is likely to be around the beginning of its Sedov-Taylor phase, which is a time of the utmost importance from the point of view of the maximum energy that shows in the overall spectrum of CRs contributed by the SNR. Here we calculate the maximum energy and the spectrum for four cases that differ slightly in terms of parameters, but they all return a shock speed of $\sim 5700-5800$ km/s at the present time and a SNR size of 2.8 pc, in agreement with the observed values \cite[]{Vink2022}. The four cases are summarized in Table \ref{table:t1}, where the values of $\zeta$ and $s$ only depend on the choice of the ejecta profile index $k$, and are provided by \cite{TangChevalier}. Case 4 was included here to account for the possibility that Cas A may have been particularly energetic: \cite{Lee2014} estimated $E_{51}=5$, $\dot M_{-5}=5$ and $M_{ej}=4 M_\odot$, with an estimated uncertainty of $\sim 50\%$. With the nominal values of these parameters, the current shock speed would exceed the value measured by \cite{Vink2022}, hence we adopted here $E_{51}=2.4$ for Case 4 (see Table \ref{table:t1}). 

\begin{figure*}[t!]
\centering
\includegraphics[width=0.49\linewidth]{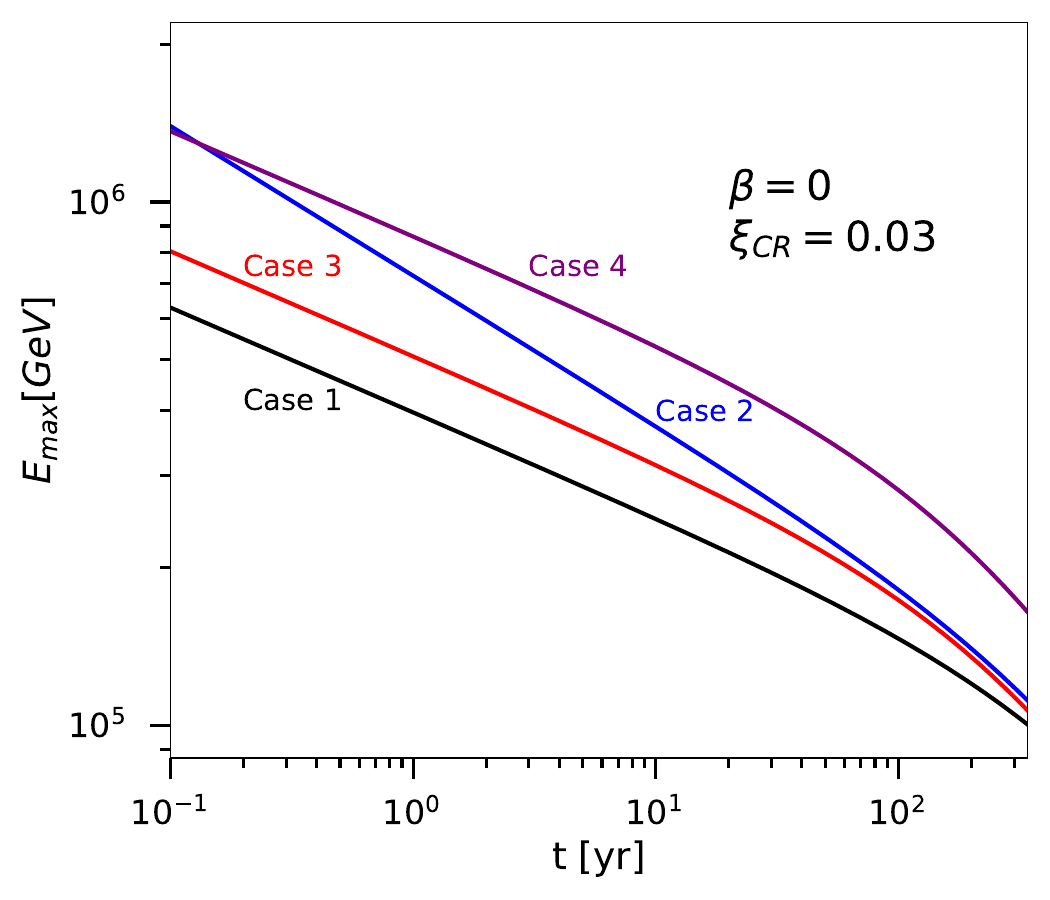}
\includegraphics[width=0.49\linewidth]{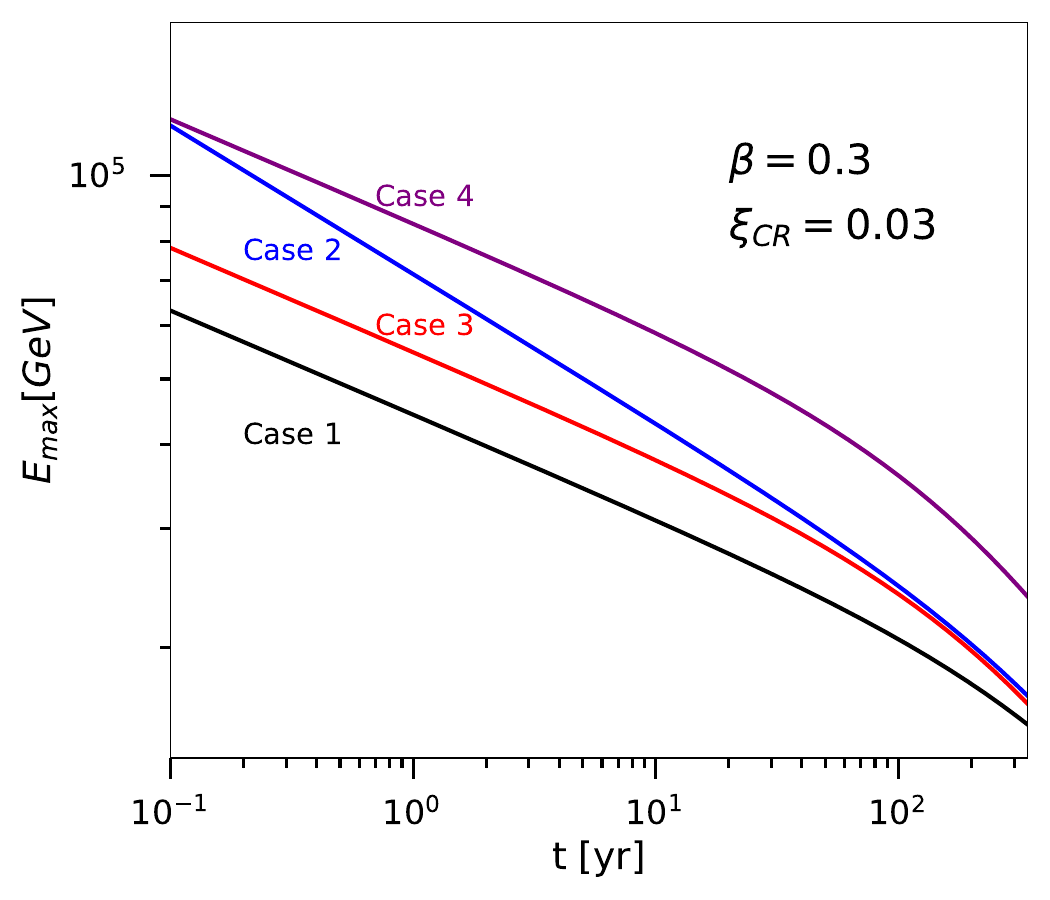}
\caption{\textbf{Left}: Maximum energy at the forward shock of Cas A in the benchmark cases discussed in the text, assuming $\beta=0$ and an acceleration efficiency of $\xi_{CR}=0.03$. \textbf{Right}: Same as the left panel but for $\beta=0.3$. The two panels illustrate the strong effect of the spectrum of accelerated particles on the maximum energy of the same particles.}
\label{fig:Emax}
\end{figure*}

In Figure \ref{fig:Emax} we show the maximum energy as a function of time for these four cases, for an acceleration efficiency $\xi_{CR}=0.03$ (in terms of pressure), and $\beta=0$ (left panel) and $\beta=0.3$ (right panel). The plots extend up to the current age of the Cas A SNR, meaning that only particles with $E_{max}$ in these plots may have been released into the ISM by Cas A. 

\begin{table}[h!]
\centering
\begin{tabular}{|c|c|c|c|c|c|c|c|}
\hline
Model & $E_{51}$ & $\frac{M_{ej}}{M_\odot}$ & $v_6$ & $\dot M_{-5}$ & $k$ & $\zeta$ & $s$\\
\hline
Case 1 & 1.1 & 3 & 2 & 1.5 & 12 & 1.14 & 3.81       \\
Case 2 & 1.1 & 2 & 2 & 2   & 9  & 0.97 & 5.16       \\
Case 3 & 1.0 & 2 & 1 & 1   & 12 & 1.14 & 3.81  \\
Case 4 & 2.4 & 4 & 2 & 5   & 12 & 1.14 & 3.81  \\
\hline
\end{tabular}
\caption{Values of parameters for the four cases discussed in the text.}
\label{table:t1}
\end{table}

\begin{figure*}[t!]
\centering
\includegraphics[width=0.49\linewidth]{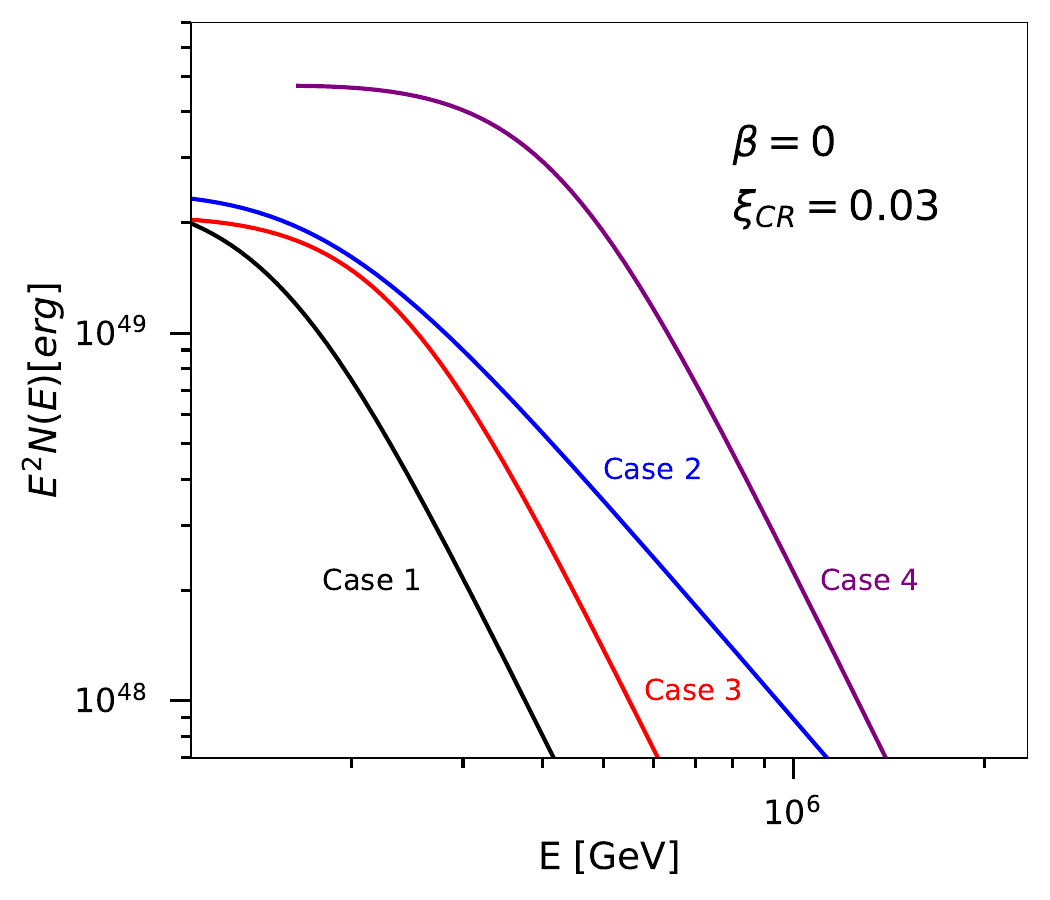}
\includegraphics[width=0.49\linewidth]{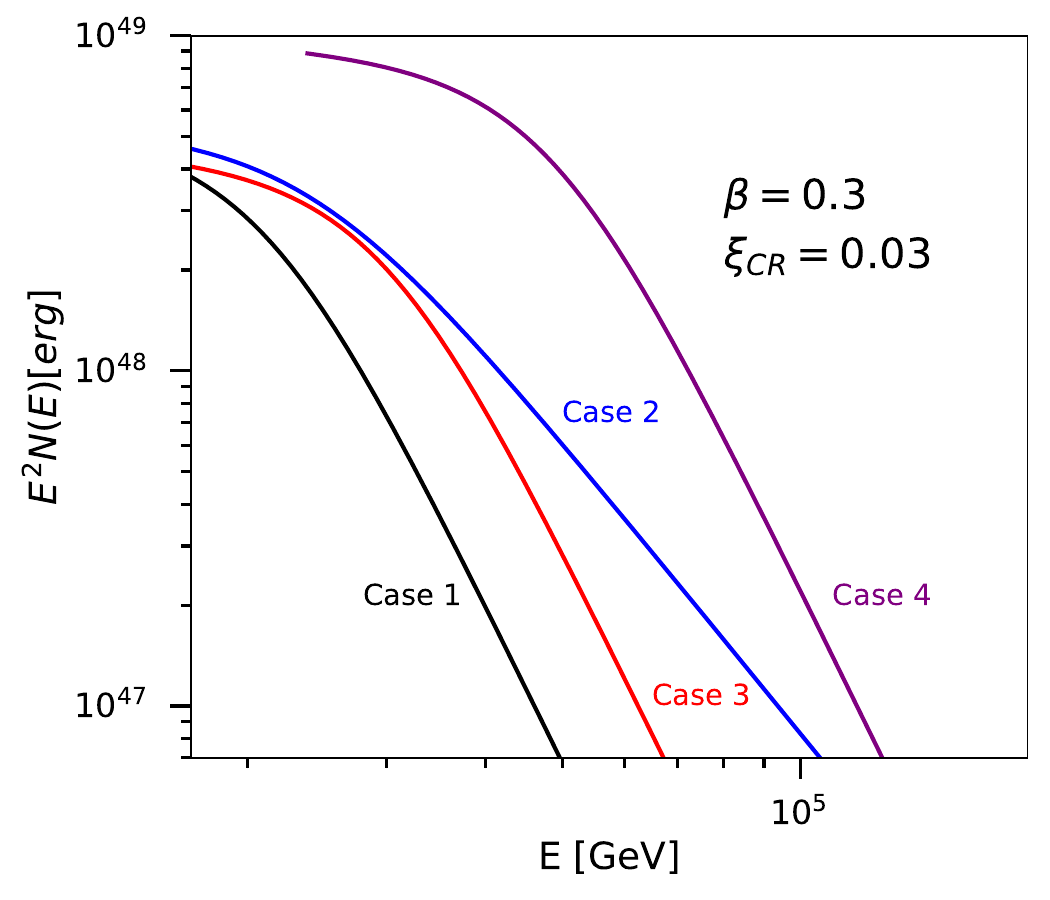}
\caption{\textbf{Left}: Spectrum of particles escaped from the Cas A SNR up until the present time, for $\beta=0$ and $\xi_{CR}=0.03$. \textbf{Right}: Same as the left panel but for $\beta=0.3$. The two panels show that for the age of Cas A, most particles have escaped during the ejecta dominated phase and they are characterized by a steep spectrum.}
\label{fig:Spectra}
\end{figure*}

Several comments are in order: 1) at early times, it is expected that Cas A indeed might have behaved as a PeVatron, in the sense of accelerating particles up to Peta-electronVolt energies. 2) The maximum energy drops in time during the ejecta dominated phase and it is currently $\sim 100$ TeV ($\sim 20$ TeV) for $\beta=0$ ($\beta=0.3$). This difference reflects the fact that larger values of $\beta$ correspond to smaller currents of escaping particles and correspondingly lower values of $E_{max}$. 3) The change of slope in $E_{max}(t)$ as a function of time reflects the corresponding change of time dependence of $R_s(t)$ and $v_s(t)$ across the transition from ejecta dominated to Sedov-Taylor phase. The spectrum of particles that have escaped Cas A until the current time can be calculated using Eq. \ref{eq:spectrum}, and is plotted in Figure \ref{fig:Spectra} (multiplied by $E^2$) for the same cases as in Figure \ref{fig:Emax}.

It may be useful to remind the reader about the general expectation: during the ejecta dominated phase the shock expands as $R_s(t)\propto t^{\frac{k-3}{k-2}}$ ($v_s\propto t^{\frac{2}{2-k}}$), while during the Sedov-Taylor phase $R_s\propto t^{2/3}$ ($v_s\propto t^{-1/3}$). Using the expressions above, one immediately concludes that the spectrum of escaping particles is the same as at the shock, $N(E)\propto E^{-2-\beta}$ during the Sedov phase, while the spectrum released during the ejecta dominated phase is much steeper, $N(E)\propto E^{-4-3\beta}$ ($N(E)\propto E^{-11/2-9\beta/2}$) for $k=9$ ($k=12$). This immediately tells us that even if particles can be accelerated up to very high energies during the early ejecta dominated phase, their spectrum at release is extremely steep, though not  exponentially suppressed. This finding reflects the fact that during such early phases very small mass is processed by the shock while propagating in the wind, despite the large wind density. 

These phenomena are all well described in Figure \ref{fig:Spectra}: since Cas A is currently at the transition time from the ejecta dominated to the Sedov-Taylor phase, the spectrum of the escaped particles is overall very steep, with the exception of the lower energy part. There are virtually no escaped particles at $E\lesssim 100$ TeV ($E\lesssim 20$ TeV) for the case $\beta=0$ ($\beta=0.3$), since this is the current maximum energy of accelerated particles. These considerations also apply to Case 4, where the energetics of the SN is somewhat larger. Moreover, for cases 1, 3, and 4, requiring $k=12$, as expected, the high energy spectrum is steeper than in case 2 ($k=9$). Based on the considerations above, at a later time, when Cas A will be deeply inside the Sedov phase, the spectrum of escaped particles will extend toward lower $E$ with a shape that reflects the spectrum at the shock, $\propto E^{-2-\beta}$. 

Figure \ref{fig:Spectra} clearly illustrates the fact that for the parameters that suit the temporal evolution of Cas A and are compatible with the current measured shock location and velocity, Cas A cannot be considered as a PeVatron. The maximum energy of accelerated particles was most likely much higher in the past, but the spectrum of such particles as they are released into the ISM is much steeper than $E^{-2}$. Obviously these findings are crucial to assess the gamma ray emission that can be expected from the interaction of escaped CRs with the environment around Cas A.

\section{Particle transport around a young SNR}
\label{sec:transport}

The recent LHAASO observation of the region around Cas A \cite[]{Cao2024} searched for diffuse emission due to CRs escaped from this young remnant and interacting with gas within a region of $\sim 2$ degrees from the SNR. At the distance of Cas A this corresponds to about 120 pc. Inside this region, \cite{Cao2024} estimated that the mean gas density is $\sim 10~\rm cm^{-3}$, due to the presence of dense molecular gas. It is clear from the $^{12}$CO line observations \cite[]{Ma2019} that this gas is distributed in a very inhomogeneous way, and more specifically it is concentrated in three main molecular structures. This is important because the rate of gamma ray production due to CR interactions with the gas is sensitive to the spatial distribution of both CRs and gas. The mean gas density in the region of interest of LHAASO is the relevant quantity to use only if CRs can ergodically probe the same volume within a time comparable with the age of the SNR. This is the issue we address below.

Based on the description provided above, the particles with the highest energies that have been released by Cas A into the surrounding medium are the ones that have been produced first, and as such they are the ones that traveled the farthest. The largest distance that they could possibly reach is $\sim c T_{SN} = 100$ pc, smaller than the region of interest (particles that propagate at an angle with respect to the local magnetic field will cover an even smaller distance). Hence all particles ever released by Cas A into the ISM are still inside the region where LHAASO carried out its observation. While this simplifies the calculation of the gamma ray emission, since no information on the spatial distribution is required, beside the spectrum $N(E)$ calculated in Section \ref{sec:escape}, it makes even more pressing the issue of what mean gas density to adopt in the calculations of the gamma ray emission. 

Let us start with the simplest scenario, in which all particles released by Cas A into the ISM move ballistically. Since the Larmor radius in the typical ISM magnetic field is $r_L\approx 0.1~\rm pc ~ E(100TeV) B_\mu^{-1}$ (where $B_\mu=(B/\mu G)$), it is clear that particles propagate along the local magnetic field lines filling a tube with an approximate cross section comparable with the size of the SNR and length $\sim c T_{SN}$. This region comprises $\sim 3.4\times 10^{-4}$ of the volume corresponding to the region of interest of LHAASO. Hence the approximation that CRs probe the whole region ergodically appears unjustified and the mean density to be used in the calculations of the gamma ray emission is not necessarily the mean density quoted by \cite{Cao2024}. Notice that the estimate above returns the largest filling factor of the volume filled with escaped CRs: any diffusive motion would force the particles to cover a smaller distance along the local field lines while the cross section of the flux tube would still be of order the size of the SNR. 

\begin{figure*}[t!]
\centering
\includegraphics[width=0.49\linewidth]{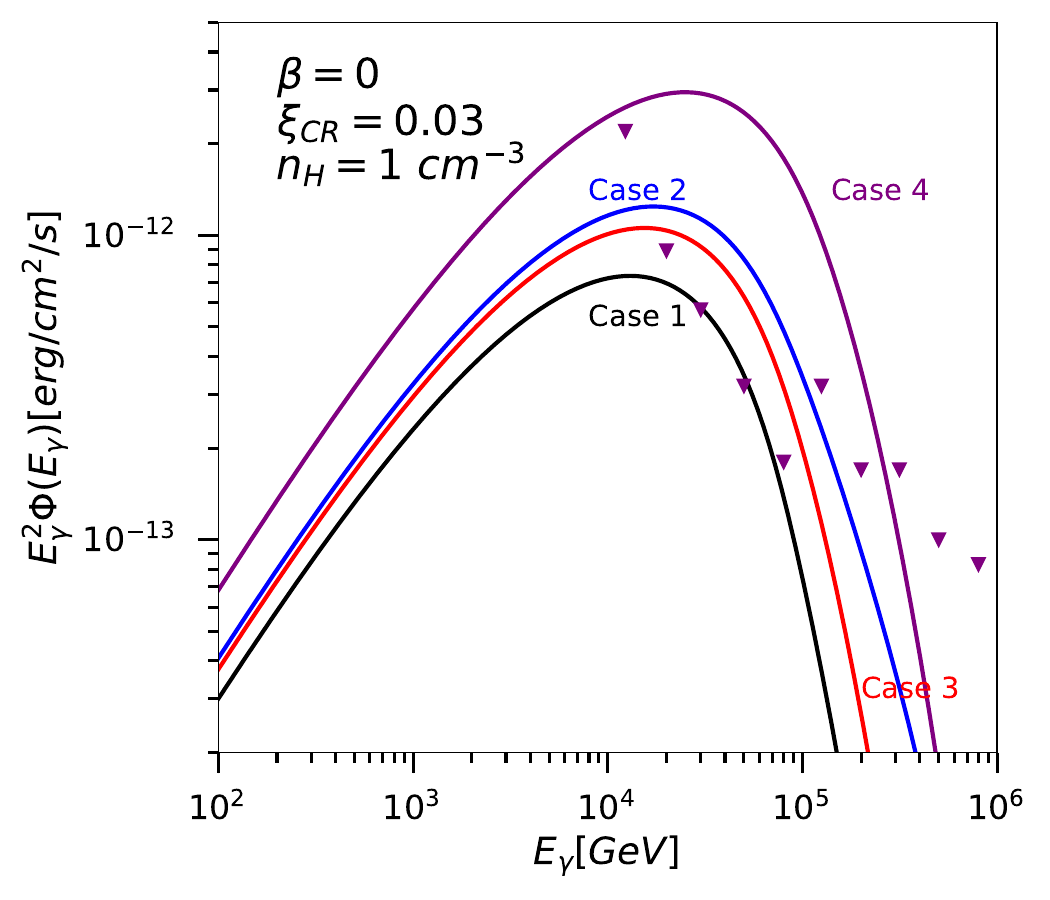}
\includegraphics[width=0.49\linewidth]{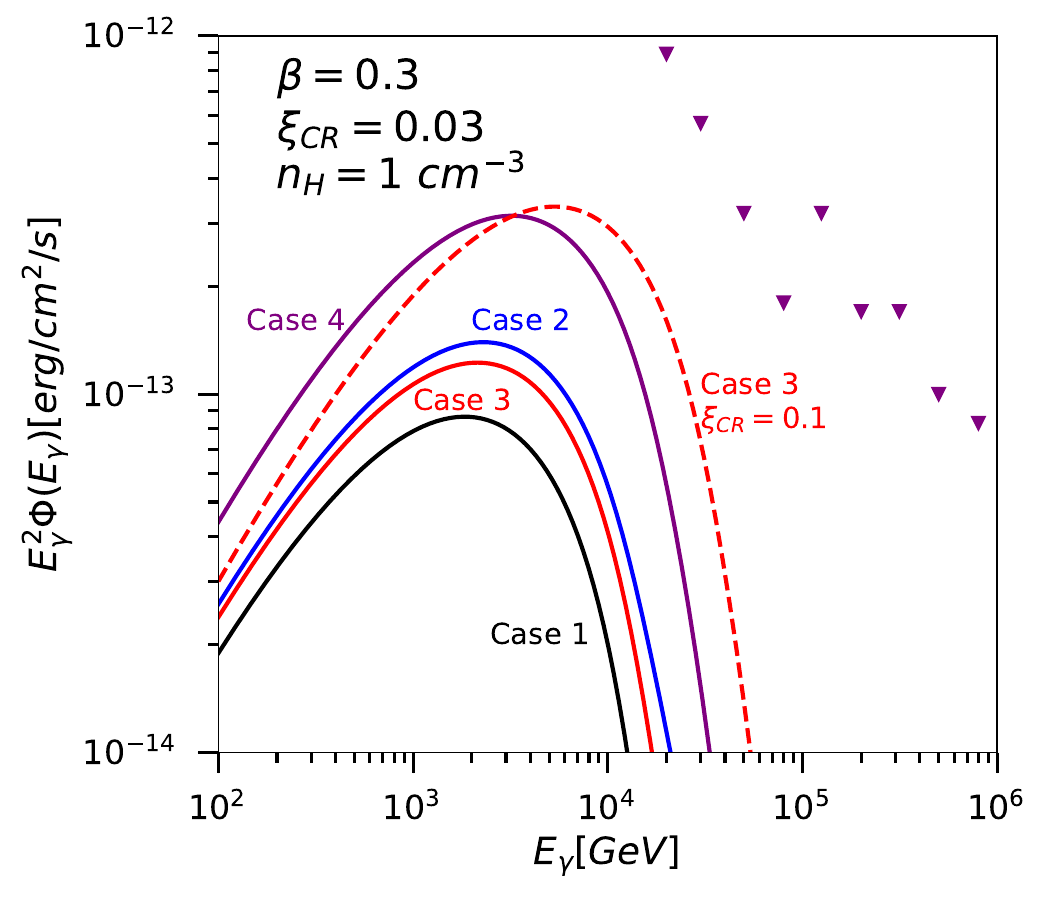}
\caption{\textbf{Left}: Gamma ray spectrum due to particles escaped from the Cas A SNR up until the present time, for $\beta=0$ and $\xi_{CR}=0.03$. \textbf{Right}: Same as the left panel but for $\beta=0.3$. The case $\beta=0.3$ and $\xi_{CR}=0.1$ is also plotted (dashed red line). The predicted gamma ray spectra are compared with the upper limits from LHAASO observation of a $2$ degrees region around Cas A \cite[]{Cao2024} (purple downward triangles).}
\label{fig:Gamma}
\end{figure*}

It is nevertheless useful to put forward some considerations concerning a possible diffusive motion: if the diffusion coefficient in the region around Cas A is similar to the one inferred from secondary/primary ratios in the Galaxy \cite[]{Evoli1}, then at rigidity $\gtrsim 300$ GV one can write $D(E)\approx 4\times 10^{28} E(GeV)^{0.4}~\rm cm^2/s$, which corresponds to a pathlength for diffusion $\lambda_D(E)\approx 120~\rm (E/100~TeV)^{0.4}~pc$. The fact that for the energies of our concern the pathlength is comparable in size with the region of interest suggests such particles are not diffusing yet. Hence the assumption of ballistic propagation is meaningful on these scales. It is worth stressing that this concept of particles leaking out of a source along local filed lines, that is often expressed in terms of anisotropic diffusion, has profound implications on the gamma ray morphology, on the longer time scales on which diffusion starts being effective \cite[]{Giacinti2013,Giacinti2022}.

One could argue that particles escaping the SNR should excite both resonant \cite[]{Nava,Dangelo,Malkov,Plesser,Bao2024} and non-resonant \cite[]{Schroer2021,Schroer2022} streaming instabilities, thereby reducing the diffusion coefficient. However, the effect of the resonant instability is typically limited to CRs with energy $\lesssim 1$ TeV \cite[]{Dangelo,Bao2024}, hence for the energies of interest here, such processes may be neglected. Non-resonant streaming instability can be effectively excited around young SNRs and grows at rate $\gamma_{CR}$ (see Section \ref{sec:escape}). We recall that by definition of maximum energy at time $t$, one has $\gamma_{CR} t \approx 5$. It follows that before the CR current is disrupted the particles have traveled quasi-ballistically for a distance $\sim 5c\gamma_{CR}^{-1}=ct$. This is due to the fact that the instability starts by exciting small scales, much smaller than the gyration radius of the particles. When the instability saturates the scale of the perturbations that get excited is comparable with the Larmor radius of the particles dominating the current calculated in the amplified magnetic field. This is the time when scattering becomes effective and the current is disrupted. In any case the particles that leave the SNR are bound to move only out to a distance $\sim ct$ and by doing so provide the conditions for particles released at later times to experience a smaller diffusion coefficient. 

In all cases considered above the region filled with escaped particles is smaller than the region of interest corresponding to $\sim 2$ degrees around Cas A, and the mean density can be much smaller than the one inferred and used by \cite{Cao2024}. An exception to this conclusion would be if one of the dense molecular clouds is magnetically connected with the SNR itself.

\section{Gamma rays from escaping particles}
\label{sec:gamma}

As illustrated in Figure \ref{fig:Spectra}, the escape process works as a high-pass filter, in that at a given time $t$ (age of the SNR), only particles with energy $\gtrsim E_{max}(t)$ may have left the remnant and populated the surrounding medium. From the point of view of gamma ray emission due to inelastic $pp$ collisions, this fact is of the utmost importance: gamma rays at energy $E_\gamma\lesssim 0.1 E_{max}(t_{SN})$ have a spectrum that reflects only the differential cross section for pion production, $\propto 1/E_\gamma$, independent of the spectrum of protons at higher energies. 

Here we discuss the calculation of gamma ray emission from the region around Cas A: as discussed above, the only relevant ingredients of this calculation are the spectrum of escaped particles $N(E)$ and the mean gas density in the region where CRs propagate. Since all particles that ever escaped Cas A are fully contained inside the region of interest used by LHAASO, the details of spatial transport discussed in Section \ref{sec:transport} have no effect on the gamma ray emission. Based on the discussion in Section \ref{sec:transport}, the mean density to be used depends on whether the SNR is magnetically connected with one of the clouds in the region. The probability for this occurrence is however rather small, therefore for the sake of definiteness we will use $n_{gas}=1~\rm cm^{-3}$ as a benchmark gas density instead of the mean density $10~\rm cm^{-3}$ in the $\sim 120$ pc around Cas A used by \cite{Cao2024}. In principle one might include in these considerations the gas that is evacuated from the near source region due to the wind of the pre-supernova star. Using Eq. \ref{eq:cavity}, we estimated that this gas should form a ring around Cas A at a distance $R_b$ of a few pc and a density of a few particles/$\rm cm^{3}$. However, unless CRs are trapped in such a ring, its effect on the gamma ray emission is not expected to be prominent given its small volume filling factor $\sim 3 (R_{b}/R_{LHAASO})^3 (\Delta R_{b}/R_{b}) \sim (0.2-2)\times 10^{-5}$, where $R_{b}\sim 2-5$ pc (see Eq. \ref{eq:cavity}), and we adopted $\Delta R_{b}/R_{b}\sim 0.1$ for the thickness of the ring in units of $R_{b}$; finally, $R_{LHAASO}=120$ pc is the size of the region of interest for LHAASO.

The flux of gamma rays, computed using the cross sections of \cite{kelner2006} (see Appendix \ref{app:A1} for more details), is shown in Figure \ref{fig:Gamma}, for the four benchmark cases discussed above and for $\beta=0$ (left panel) and $\beta=0.3$ (right panel). The purple triangles are the upper limits reported by LHAASO \cite[]{Cao2024}. 

In the case $\beta=0$ that maximizes the current of escaping CRs and hence leads to higher values of the maximum energy, the expected gamma ray fluxes are at the same level as the upper limits, with cases 2 and 3 being in mild tension with them and Case 4 exceeding the upper limits by about a factor $\sim 2-3$ between 10 and 100 TeV. The spectrum at $E_\gamma\lesssim 10$ TeV is $\sim 1/E_\gamma$, as expected based on the energy dependence of the differential cross section for gamma ray production, since there are no CRs at low energies that can directly contribute such gamma rays. Notice that at least in cases 2 and 4 the maximum energy in the early stages can be as high as $\sim$Peta-electronVolt (see Figure \ref{fig:Emax}), and yet the spectrum of such high energy-particles is very steep and does not lead to an appreciable gamma ray emission at the present time. 

In the case $\beta=0.3$ (steeper spectrum at the shock, possibly induced by the formation of a postcursor \cite[]{Caprioli+2020}), the current of escaping particles is smaller and the maximum energy correspondingly lower. The gamma ray fluxes (right panel) are all much lower than the LHAASO upper limits. In fact they remain compatible with such limits even for a larger gas density and/or a larger value of the CR acceleration efficiency (see for instance the dashed line for case 3 and $\xi_{CR}=0.1$). 

Figure \ref{fig:Gamma} shows a few points very clearly: despite Cas A being close to the beginning of its Sedov phase from the evolutionary point of view, its maximum energy is well below $\sim$Peta-electroVolt. This is however perfectly in line with the standard theory of DSA at non relativistic shocks with magnetic field amplification. Invoking additional processes that may push the expected maximum energy for Cas A toward larger values appears at this time not supported by any evidence of the existence of such particles. The maximum energy of particles accelerated in Cas A has certainly been higher at earlier times but these particles leave little imprint in the spectrum of CRs released into the ISM, $N(E)$, as discussed in Section \ref{sec:escape}: the effective maximum energy manifests itself in the form of a steepening from a power law $\sim E^{-2-\beta}$ (extending down to the maximum energy at the end of the Sedov-Taylor phase, when that phase will be eventually reached) to a steeper one (depending on the density profile of the ejecta) at an energy that corresponds to the highest energy of accelerated particles at the beginning of the Sedov-Taylor phase. This finding is also reflected in the gamma ray emission. In the perspective of assessing the role of a SNR as a potential PeVatron, this is the maximum energy that counts. 

\section{Discussion}
\label{sec:discuss}

The debate on which CRs can possibly originate in SNRs has now been ongoing for four decades: while there is little doubt that the bulk of the low energy CRs are actually accelerated at the forward shock of SNRs, there is still quite some controversy on whether these sources can account for CRs in the knee region. The debate is mainly articulated in two parts. On one hand, there is an observational argument: no SNR has ever been observed to be accelerating CRs with energy $\gtrsim 100$ TeV. While this is a serious point, one could argue that perhaps the SNRs that have been observed in gamma rays are not in the stage of their evolution in which the highest energies are reached: perhaps the SNR has operated as a PeVatron at some earlier time. 

Then there is a more theoretical point: it became clear early on \cite[]{Bell:1978p1342,Bell:1978p1344,Lagage1,Lagage2} that DSA could lead to reasonably high energies of accelerated particles only in the presence of streaming instability, excited by the same accelerated particles. Even with this noteworthy addition, the maximum energy was predicted to be $\lesssim 10-100$ TeV \cite[]{Lagage1,Lagage2}. After the discovery of a non-resonant branch of the instability \cite[]{Bell:2004p737}, growing much faster than the resonant one, the situation became more optimistic about SNRs reaching Peta-electronVolt energies. However, this instability appears to be very important for extremely young SNRs and when the shock propagates in the dense wind of the pre-supernova star (core collapse supernovae), while the maximum energy at the beginning of the Sedov-Taylor phase still falls short of Peta-electronVolt by some factor, depending on the properties of the SNR \cite[]{Schure-Bell:2013,pierre}. While this conclusion is not at odds with any piece of observation, it leaves open the possibility that other instabilities (see for instance \cite[]{Beresnyak2009,Drury2012}) or other processes (see for instance \cite{Bell2025} on the role of mirroring) could operate to boost the maximum energy of accelerated particles.  

The Cas A SNR is the perfect target of gamma ray observations to address the issues above: first, Cas A is of the right type to reach very high maximum energies, being a type IIb SNR with the shock expanding in the wind of a red giant progenitor. Moreover Cas A is expected to be at the transition from late ejecta dominated phase to early Sedov-Taylor phase, a crucial stage for particle acceleration. Finally, the LHAASO observation of the region around Cas A gives us the opportunity to investigate not only if very high- energy particles are being accelerated at present but also if they were produced in the past of the SNR evolution. 

Here we investigated these points using the ongoing observation of the dynamical evolution of the Cas A forward shock \cite[]{Vink2022}, and the LHAASO upper limits on gamma ray emission from a region of $\sim 2$ degrees around Cas A \cite[]{Cao2024} as guidance. For the parameters of Cas A that make its evolution compatible with the current state of the SNR, the maximum energy ranges between $\sim 20$ TeV (for $\beta=0.3$) and $\sim 100$ TeV (for $\beta=0$), well short of the knee region. This implies that the current theoretical understanding of particle acceleration, based on the development of a non-resonant streaming instability, do not make Cas A a PeVatron candidate at present. Nevertheless, the same theoretical framework leads to the expectation that, at least for some models of evolution, Cas A should have accelerated particles to energies $\sim 1$ PeV in early phases.  

We showed that the spectrum of the particles escaped in the past epochs of Cas A is very steep (typically $\propto E^{-4}-E^{-5}$), reflecting the crucial fact that, from the point of view of the SNR contribution to the flux of CRs, the main contribution comes from the Sedov-Taylor phase, that Cas A is only entering now. The non-thermal particles that have escaped the Cas A SNR are all inside the region of interest observed by LHAASO (about $\sim 120$ pc wide at the distance of 3.4 kpc). Hence the gamma ray emission produced by the interactions of the escaped CRs with the environment only depends on the spectrum of particles in the region, $N(E)$ (resulting from the integration over the past ages of the SNR evolution), and the mean density in the region reached by CRs. The latter was estimated by \cite{Cao2024} to be $\sim 10\,\rm cm^{-3}$, mainly based on a few dense massive clouds in the region. We argued that this would be the correct density to use for the gamma ray production only if CRs could probe the whole region, perhaps through diffusion. This is not the case: on the time scales of the age of the SNR, only a small volume is probed by CRs, shaped by the size of the SNR (2.8 pc) and the local magnetic field lines, along which particles are likely moving ballistically. If one of the clouds were magnetically connected to the SNR, it would be illuminated directly by the escaping particles, but the probability for this to happen is very small. Hence, smaller values of the mean density are more easily justified (we used $\sim 1\,\rm cm^{-3}$ as a benchmark value).

The best hints to the spectrum of CRs accelerated at the forward shock at the present time is provided by the lower energy gamma ray observation of the remnant itself, which suggests a spectrum slightly softer than $E^{-2}$ (namely $\beta>0$) and a maximum energy of accelerated particles $\sim 20$ TeV \cite[]{Magic2017,Veritas2020,Cao2025}. The gamma ray emission expected from a region of $\sim 2$ degrees around Cas A due to escaped CRs is marginally compatible with the LHAASO upper limits from the same region for $\beta=0$, while it is well below the upper limits for $\beta=0.3$.

It is perhaps worth stressing that neither the gamma ray observations of the SNR nor the upper limits from the region around it require that the maximum energy of accelerated particles is different from the one inferred based on the theory of DSA with non-resonant streaming instability as a source of magnetic field amplification. We can conclude that Cas A is unlikely to have contributed in any appreciable way to the Peta-electronVolt CR flux observed at Earth. Although it is difficult to generalize the results obtained on an individual SNR to a class of objects, we can certainly say that at present there are strong indications that if some SNRs act as real PeVatrons, meaning that the maximum energy at the beginning of the Sedov phase is in the Peta-electronVolt range, then these must be associated with rare and possibly very energetic supernova explosions \cite[]{pierre,pierre2021}. A potential caveat to these conclusions might be introduced by the assumption that the SNR shock propagates in a smooth density profile of the wind of the SN progenitor: some evidence suggests that $\dot M$ may increase with time \cite[]{Beasor2018}, leading to larger mass loss rates in the late stages of the red giant life. A reliable model of these phenomena, especially in the case of Cas A, is however not available at this time, and it is therefore rather difficult to assess their importance for the particles escaping a SNR. Qualitatively, a larger $\dot M$ in the late stages of the red giant evolution implies that the SNR shock in the beginning moves in a region with larger density, and hence larger maximum energies may be achieved. It remains true, however, that the spectrum of these particles, as they are released into the ISM, is very steep, unless the SNR enters its Sedov phase so early on, which is unlikely to happen, in that it would imply unreasonably high values of the mass loss rate.  

\begin{acknowledgements}
The author is grateful to E. Amato, A. Capanema, P. Cristofari and E. Sobacchi for providing useful comments on the manuscript. This work has been partially funded by the European Union - Next Generation EU, through PRIN-MUR 2022TJW4EJ and by the European Union - NextGenerationEU under the MUR National Innovation Ecosystem grant  ECS00000041 - VITALITY/ASTRA - CUP D13C21000430001.
\end{acknowledgements}

\bibliographystyle{aa}
\bibliography{biblio}

\begin{appendix}
\section{Calculation of gamma ray emission}
\label{app:A1}

The flux of gamma rays with energy $E_\gamma>0.1$ TeV from the region around Cas A can be written following \cite{kelner2006}, as:
\begin{equation}
    \Phi_\gamma(E_\gamma)=\frac{c n_H}{4\pi D^2} \int_0^1 \frac{dx}{x} \sigma_{in}(E_\gamma/x)F_\gamma(x,E_\gamma/x) N(E_\gamma/x),
    \label{eq:flux}
\end{equation}
where 
\begin{eqnarray}
    F_\gamma(x,E)=B_\gamma \frac{\ln x}{x}\left[\frac{1-x^{\beta_\gamma}}{1+k_\gamma x^{\beta_\gamma} (1-x^{\beta_\gamma})} \right]^4 \times\\
    \left\{ \frac{1}{\ln x} - \frac{4 \beta_\gamma x^{\beta_\gamma}}{1-x^{\beta_\gamma}} - \frac{4 k_\gamma \beta_\gamma x^{\beta_\gamma} (1-2 x^{\beta_\gamma})}{1+k_\gamma x^{\beta_\gamma}(1-x^{\beta_\gamma})} \right\},
\end{eqnarray}
D is the distance to Cas A and the coefficients are as given by \cite{kelner2006}:
\begin{eqnarray}
    B_\gamma = 1.3+ 0.14 L + 0.011 L^2, \\
    \beta_{\gamma} = \frac{1}{1.79+0.11 L + 0.008 L^2}, \\
    k_\gamma= \frac{1}{0.801 + 0.049 L + 0.014 L^2},
\end{eqnarray}
with $L=\ln(E/TeV)$ and $E$ is the proton energy. The expressions above hold as long as the proton energy is $\gtrsim 0.1$ TeV, a condition that is amply satisfied in our calculations. The inelastic cross section in Eq. \ref{eq:flux} reads:
\begin{equation}
\sigma_{in}(E)=34.3+1.88 L + 0.25 L^2~\rm mb.
\end{equation}

\end{appendix}

\end{document}